\documentclass[twocolumn, trackchanges]{aastex62} 
\usepackage{amsmath}
\usepackage{amssymb}
\usepackage{mathtools}
\usepackage{mathrsfs}
\usepackage{bm}
\usepackage{enumerate}
\usepackage{booktabs}
\graphicspath{{./}{figures/}}

\newcommand{\project}[1]{\textsl{#1}}
\usepackage{xspace}

\newcommand*{\NICER}{\project{NICER}\xspace}
\newcommand*{\MultiNest}{\textsc{MultiNest}\xspace}

\newcommand{\msol}{M$_\odot$\xspace}
\newcommand{\jdbl}{PSR~J0030$+$0451\xspace}
\newcommand{\joh}{PSR~J0740$+$6620\xspace}

\shorttitle{EOS and neutron star properties from \NICER and multimessenger observations}
\shortauthors{Raaijmakers et al.}

\begin{document}

\correspondingauthor{G.~Raaijmakers}
\email{G.Raaijmakers@uva.nl}

\author[0000-0002-9397-786X]{G.~Raaijmakers}
\affil{GRAPPA, Anton Pannekoek Institute for Astronomy and Institute of High-Energy Physics, University of Amsterdam, Science Park 904, 1098 XH Amsterdam, Netherlands}

\author[0000-0001-8641-2062]{S.~K.~Greif}
\affil{Technische Universit\"at Darmstadt, Department of Physics, 64289 Darmstadt, Germany}
\affil{ExtreMe Matter Institute EMMI, GSI Helmholtzzentrum f\"ur Schwerionenforschung GmbH, 64291 Darmstadt, Germany}

\author[0000-0003-0640-1801]{K.~Hebeler}
\affil{Technische Universit\"at Darmstadt, Department of Physics, 64289 Darmstadt, Germany}
\affil{ExtreMe Matter Institute EMMI, GSI Helmholtzzentrum f\"ur Schwerionenforschung GmbH, 64291 Darmstadt, Germany}

\author[0000-0002-3394-6105]{T.~Hinderer}
\affil{Institute for Theoretical Physics, Utrecht University, Princetonplein 5, 3584CC Utrecht, The Netherlands}

\author[0000-0001-6573-7773]{S.~Nissanke}
\affil{GRAPPA, Anton Pannekoek Institute for Astronomy and Institute of High-Energy Physics, University of Amsterdam, Science Park 904, 1098 XH Amsterdam, Netherlands}
\affil{Nikhef, Science Park 105, 1098 XG Amsterdam, The Netherlands}

\author[0000-0001-8027-4076]{A.~Schwenk}
\affil{Technische Universit\"at Darmstadt, Department of Physics, 64289 Darmstadt, Germany}
\affil{ExtreMe Matter Institute EMMI, GSI Helmholtzzentrum f\"ur Schwerionenforschung GmbH, 64291 Darmstadt, Germany}
\affil{Max-Planck-Institut f\"ur Kernphysik, Saupfercheckweg 1, 69117 Heidelberg, Germany}

\author[0000-0001-9313-0493]{T.~E.~Riley}
\affil{Anton Pannekoek Institute for Astronomy, University of Amsterdam, Science Park 904, 1090GE Amsterdam, the Netherlands}

\author[0000-0002-1009-2354]{A.~L.~Watts}
\affil{Anton Pannekoek Institute for Astronomy, University of Amsterdam, Science Park 904, 1090GE Amsterdam, the Netherlands}

\author{J.~M.~Lattimer}
\affil{Department of Physics and Astronomy, Stony Brook University, Stony Brook, NY 11794-3800, USA}

\author[0000-0002-6089-6836]{W.~C.~G.~Ho}
\affil{Department of Physics and Astronomy, Haverford College, 370 Lancaster Avenue, Haverford, PA 19041, USA}

\title{Constraints on the dense matter equation of state and neutron star properties from\\
\NICER's mass-radius estimate of \joh and multimessenger observations}

\begin{abstract}
In recent years our understanding of the dense matter equation of state (EOS) of neutron stars has significantly improved by analyzing multimessenger data from radio/X-ray pulsars, gravitational wave events, and from nuclear physics constraints. Here we study the additional impact on the EOS from the jointly estimated mass and radius of \joh, presented in \citet{Riley21} by analyzing a combined dataset from X-ray telescopes \NICER and \textit{XMM-Newton}. We employ two different high-density EOS parameterizations: a piecewise-polytropic (PP) model and a model based on the speed of sound in a neutron star (CS). At nuclear densities these are connected to microscopic calculations of neutron matter based on chiral effective field theory interactions. In addition to the new \NICER data for this heavy neutron star, we separately study constraints from the radio timing mass measurement of \joh, the gravitational wave events of binary neutron stars GW190425 and GW170817, and for the latter the associated kilonova AT2017gfo. By combining all these, and the \NICER mass-radius estimate of \jdbl, we find the radius of a $1.4\,$\msol neutron star to be constrained to the 95\% credible ranges $12.33^{+0.76}_{-0.81}\,$km (PP model) and $12.18^{+0.56}_{-0.79}\,$km (CS model). In addition, we explore different chiral effective field theory calculations and show that the new \NICER results provide tight constraints for the pressure of neutron star matter at around twice saturation density, which shows the power of these observations to constrain dense matter interactions at intermediate densities.
\end{abstract}


\section{Introduction}
\label{sec:intro}

Our understanding of the dense matter equation of state (EOS) of neutron stars has made significant progress over the last few years due to the arrival of new avenues to measure observables like mass, radius and tidal deformability, that connect to the behavior of matter at supranuclear densities. Recently NASA's X-ray timing telescope, the \textit{Neutron Star Interior Composition Explorer} (\NICER), has delivered the first joint measurement of mass and radius through pulse profile modeling of the millisecond pulsar PSR~J0030$+$0451 \citep[][]{Riley19, Miller19}. The impact of this measurement on the dense matter EOS has been extensively studied in various EOS frameworks \citep[see, e.g.,][]{Raaijmakers19,Miller19,Raaijmakers20,Essick20,Landry20,Dietrich20,Jiang20,AlMamun21}, including EOS with phase transitions to quark matter \citep[see, e.g.,][]{XieWJ20c, LiA20,Tang20,Blaschke20,Alvarez20b} and models that explore the possibility of there being two stable neutron star branches \citep{Christian20}. 

Concurrently, the second and third observing runs of LIGO/Virgo have so far resulted in the confirmed gravitational wave detections of two (most-likely) binary neutron star mergers: GW170817 \citep{GW170817discovery, GW170817sourceproperties} and GW190425 \citep{GW190425}. By accurately measuring the gravitational wave phase, limits can be put on the EOS-dependent tidal deformability of the neutron stars \citep{Flanagan:2007ix, Hinderer10}. While for GW170817 the tidal deformability could be measured within a $90\%$ highest posterior density interval when adopting low spin priors \citep[see, e.g.][]{Abbott18,LIGO_161218_catalog}, the low signal-to-noise ratio (SNR) of GW190425 resulted in only weak
upper limits on the tidal deformability even when assuming low spins \citep{GW190425}. We consider the $\sim2.6$ \msol secondary object in GW190814 \citep{Abbott20_GW190814} to be a black hole \citep{Nathanail21}, and will therefore not use this third event in our analysis.

At nuclear densities, the EOS is well constrained by nuclear theory and experiments \citep[see, e.g.,][]{Tsan12esymm,Latt12esymm,Huth21}. In particular, many-body calculations based on chiral effective field theory (EFT) interactions have enabled systematic predictions for the neutron matter EOS up to nuclear saturation density including theoretical uncertainties \citep[see, e.g.,][]{Hebeler13,Tews13,Lynn16,Drischler19,Drischler20}. Up to saturation density, the resulting symmetry energy and pressure of neutron matter are also consistent with extractions from nuclear experiments \citep{Latt12esymm}, including from measurements of the dipole polarizability of neutron-rich nuclei \citep{Roca-Maza15,Birkhan17,Kaufmann2020}. Taking these results at nuclear densities, combined with standard crust EOS, different extrapolations to high densities have been found to lead to NS radii consistent with all multimessenger observations \citep[see, e.g.,][]{Raaijmakers20,Essick20,Annala20,Dietrich20,Biswas20}. Recently, the results of PREX-II have pointed to higher pressures \citep{Adhikari21,Reed21}, but with very large uncertainties, so that in a combined analysis with astrophysical and chiral EFT constraints, the overall consistency still persists \citep{Essick21}.

\NICER data has now enabled a joint estimate of the mass and radius of the high-mass rotation-powered millisecond pulsar \joh. Since \joh (unlike \jdbl) is in a binary with an inclination that allows measurement of the Shapiro delay, its mass can be measured independently via radio timing.  \citet{Cromartie19} reported a mass of $2.14^{+0.10}_{-0.09}$ \msol, and a joint campaign by  the \project{North American Nanohertz Observatory for Gravitational Waves} (NANOGrav) and the \project{Canadian Hydrogen Intensity Mapping Experiment} (CHIME)/Pulsar collaborations has now resulted in an updated mass of $2.08\pm 0.07$ \msol \citep{Fonseca21}.  

\citet{Riley21} have used this mass measurement as an informative prior for pulse-profile modeling analysis that is joint over the phase-resolved spectroscopic data from \NICER and phase-averaged data from the {\it XMM-Newton} European Photon Imaging Camera (EPIC). The inclusion of the smaller {\it XMM-Newton} (hereafter \textit{XMM}) data set allows for better constraints on the proportion of the X-ray emission that is attributable to background rather than \joh, ultimately acting to cut out solutions with high compactness. This results in an inferred radius of $12.39_{-0.98}^{+1.30}$\,km, and a mass of  $2.072_{-0.066}^{+0.067}$\,M$_{\odot}$ that is little changed from the radio prior.  For a full description of the methodology employed in the mass-radius inference we refer the reader to \citet{Riley21}.  

In this Letter, we use the mass and radius from \citet{Riley21} for \joh as input for inferring the dense matter EOS, combining it with other constraints from nuclear theory and multi-messenger observations. It should be considered as a follow-up to our previous work that built on \NICER's results for \jdbl \citep{Raaijmakers19,Raaijmakers20}, where in this work we explore also a broader range of multi-messenger constraints. As the high-density constraints from astrophysical observations get more precise, with the new \NICER results and future LIGO/Virgo measurements, it will be intriguing to see them play out with the present nuclear constraints. In this Letter, we also explore this for the new \NICER results and how they constrain the EOS above nuclear densities starting from different chiral EFT calculations\footnote{The posterior samples and scripts to make the plots in this Letter are available in a Zenodo repository at \citet{plotdata2}.}.

\section{Inference framework}

In this work we will closely follow the analysis framework developed previously in \citet{Greif19}, \citet{Raaijmakers19} and \citet{Raaijmakers20}. Below, we summarize this method and highlight several updates to the framework.

We consider two EOS parameterizations: i) a piecewise polytropic (PP) model with three segments between varying transitions densities \citep[][]{Hebeler13}, and ii) a speed-of-sound (CS) model first introduced in \citet{Greif19}. To capture the uncertainty in the EOS around nuclear saturation density ($n_0=0.16 \, \text{fm}^{-3}$), both parameterizations are matched to a power law fit of a range of EOS calculated from chiral effective field theory interactions \citep{Hebeler10a, Hebeler13} below $1.1 n_0$. At densities below $0.5n_0$ this power law fit is connected to the BPS crust EOS \citep{Baym71}.

\begin{figure*}[t!]
\centering
\includegraphics[width=.95\textwidth]{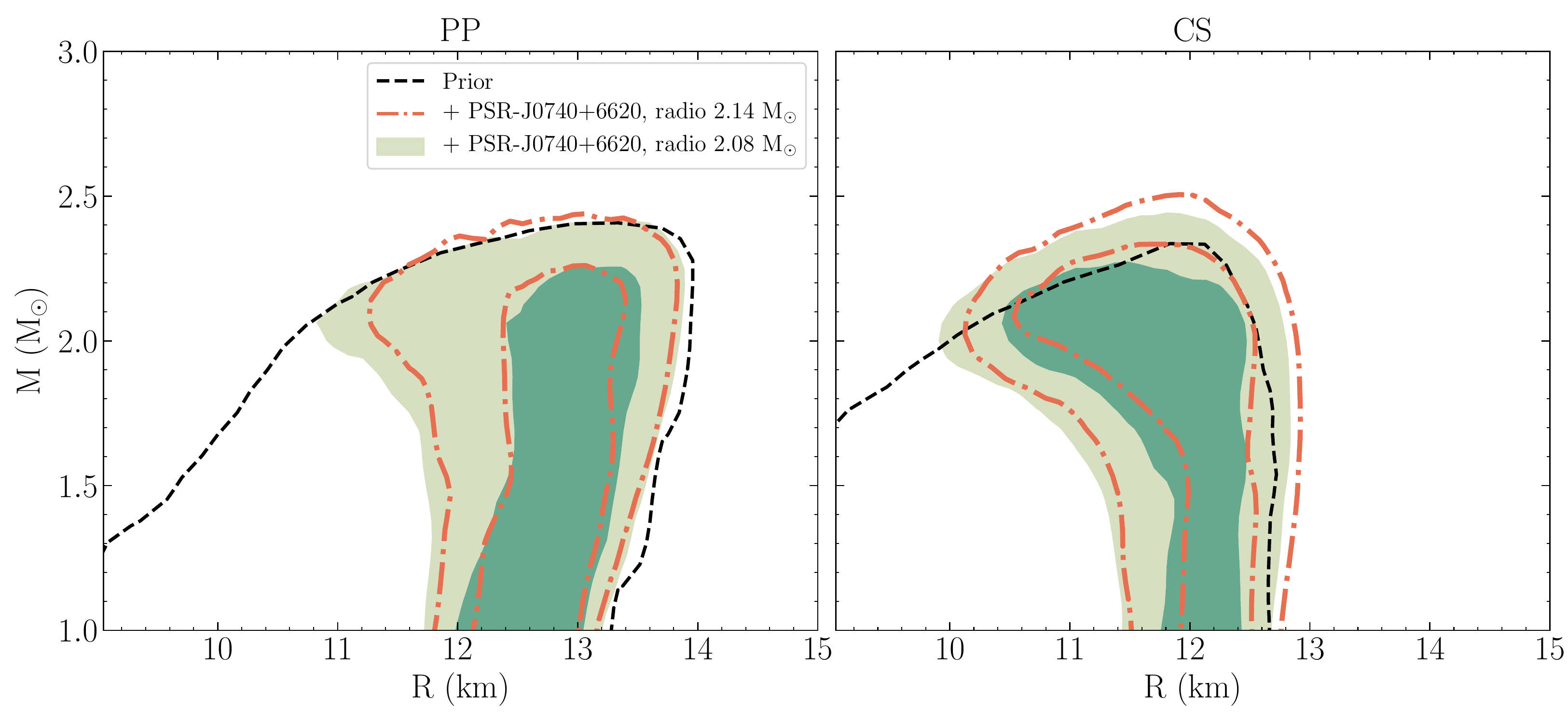}
\caption{Constraints on the mass-radius relation of neutron stars, given the posterior distribution on EOS parameters $\bm{\theta}$ using the PP model (left) and CS model (right panel). The constraints from the updated radio timing mass of \joh from \citet{Fonseca21} (present work, green) are compared to the mass from \citet{Cromartie19} used in our previous works \citep{Raaijmakers19,Raaijmakers20} (orange, dashed-dotted), showing both the $68\%$ and $95\%$ credible regions. The black dashed lines indicate the $95\%$ credible region of the prior distribution. Note that the slightly lower mass measurement does not have a significant impact on the EOS posterior.}
\label{fig:fig1}
\end{figure*}

To constrain these EOS parameterizations, governed by the EOS parameters $\bm{\theta}$, we employ Bayes' theorem and write the posterior distributions of the EOS parameters and central energy densities $\bm{\varepsilon}$ as
\begin{equation}
\label{eq:eq1}
p(\bm{\theta}, \bm{\varepsilon} \,|\, \bm{d}, \mathbb{M})
\propto 
p(\bm{\theta} \,|\, \mathbb{M})
~
p(\bm{\varepsilon} \,|\, \bm{\theta}, \mathbb{M})
~
p(\bm{d} \,|\, \bm{\theta}, \mathbb{M}) \,,
\end{equation}
where $\mathbb{M}$ denotes the model including all assumed physics and $\bm{d}$ the dataset used to constrain the EOS, consisting of, e.g., radio-, X-ray and gravitational wave data. When assuming each of these datasets to be independent of each other, we can separate the likelihoods and write 
\begin{align}
p(\bm{\theta}, \bm{\varepsilon} \,|\, &\bm{d}, \mathbb{M})
\propto 
p(\bm{\theta} \,|\, \mathbb{M})
~
p(\bm{\varepsilon} \,|\, \bm{\theta}, \mathbb{M}) \nonumber\\[1mm]
& \times \prod_{i} p(\Lambda_{1,i}, \Lambda_{2,i}, M_{1,i}, M_{2,i} \,|\, \bm{d}_{\rm GW, i}(, \bm{d}_{\rm EM, i})) \nonumber\\
& \times \prod_{j} p(M_j, R_j \,|\, \bm{d}_{\rm NICER,j}) \nonumber\\
& \times \prod_{k} p(M_k \,|\, \bm{d}_{\rm radio,k}) \,.
\label{eq:eq2}
\end{align}
Here the products run over the number of different observed stars, or mergers, in the case of the gravitational wave data. Furthermore, in Equation (\ref{eq:eq2}) we have equated the nuisance-marginalized likelihoods to the nuisance-marginalized posterior distributions derived in \citet{Riley19, Fonseca21, Riley21, GW170817sourceproperties, GW190425}. This approximation is justifiable when the priors used in estimating these nuisance-marginalized posterior distributions are uninformative, which for simplicity we will assume to be a uniform prior in this case. The posterior distributions derived by \citet{Riley19} and \citet{Riley21} already use a jointly uniform prior in mass and radius. The posterior distributions derived by \citet{GW170817sourceproperties} and \citet{GW190425} use a jointly uniform prior in the tidal deformabilities of the two components $\Lambda_i$ within the range $\Lambda_i \subset [0, 5000]$ (for GW190425 the upper bound of $\Lambda_2$ was set to $10^4$.). The prior on the detector frame masses, which are redshifted with respect to the source frame masses ($M_{\rm det} = M_{i}(1 + z)$), is uniform within the range $M_{\rm det} \subset [0.5, 7.7]$ and $M_{\rm det} \subset [1, 5.31]$ for GW170817 and GW190425 respectively. However, the posterior distribution on component masses from gravitational waves is highly degenerate because of the accurately measured chirp mass $\mathcal{M}_c = (M_1 M_2)^{3/5}/(M_1 + M_2)^{1/5}$. To speed up the convergence of our parameter estimation, we therefore transform the gravitational wave posterior distributions to include the two tidal deformabilities, chirp mass and mass ratio $q$, while reweighing such that the prior distribution on these parameters is uniform. Further, we also fix the chirp mass to its median value, since the small uncertainty in this parameter does not affect the EOS parameter estimation \citep[see][]{Raaijmakers20}, and thus have: 
\begin{align}
p(\bm{\theta}, \bm{\varepsilon} \,|\, &\bm{d}, \mathbb{M})
\propto 
p(\bm{\theta} \,|\, \mathbb{M})
~
p(\bm{\varepsilon} \,|\, \bm{\theta}, \mathbb{M}) \nonumber\\[1mm]
& \times \prod_{i} p(\Lambda_{1,i}, \Lambda_{2,i}, q_i \,|\, \mathcal{M}_c, \bm{d}_{\rm GW, i}(,\bm{d}_{\rm EM,i})) \nonumber\\
& \times \prod_{j} p(M_j, R_j \,|\, \bm{d}_{\rm NICER,j}) \nonumber\\
& \times \prod_{k} p(M_k \,|\, \bm{d}_{\rm radio,k}) \,.
\label{eq:eq3}
\end{align}
Fixing the chirp mass means that the vector $\bm{\varepsilon}$ only contains one central density per merger, where the tidal deformability of the second component is now set by $\Lambda_{2}=\Lambda_{2}(\bm{\theta};q)$. If a gravitational wave event has an associated electromagnetic (EM) counterpart, the likelihood for that event becomes a product of the nuisance-marginalized posterior distribution from the gravitational wave data and the nuisance-marginalized posterior distribution from the EM analysis, such that:
\begin{align}
p(\Lambda_{1}, \Lambda_{2}, q \,|\, \mathcal{M}_c, \bm{d}_{\rm GW}, &\bm{d}_{\rm EM}) \propto p(\Lambda_{1}, \Lambda_{2}, q \,|\, \mathcal{M}_c, \bm{d}_{\rm GW}) \nonumber \\[1mm]
& \times p(\Lambda_{1}, \Lambda_{2}, q \,|\, \mathcal{M}_c, \bm{d}_{\rm EM}) \,.
\end{align}
Obtaining the posterior distribution $p(\Lambda_{1}, \Lambda_{2}, q \,|\, \mathcal{M}_c, \bm{d}_{\rm EM})$ is discussed in Section \ref{subsec:at2017gfo} for the specific case of AT2017gfo, the kilonova associated with GW170817 \citep[see, e.g.,][]{LVCmultimessenger,GW170817grb, arcavi17, GW170817swope, Chornock17, Cowperthwaite17, Kasliwal17, Nicholl17, Tanvir17}.

We then sample from the posterior distribution $p(\bm{\theta}, \bm{\varepsilon} \,|\, \bm{d}, \mathbb{M})$, compute the corresponding $M$, $R$, and $\Lambda$, and then evaluate the likelihood by applying a kernel density estimation to the posterior distributions from \citet{Riley19, Riley21, GW170817sourceproperties, GW190425} using the nested sampling software \MultiNest. The same prior distribution $p(\bm{\theta} \,|\, \mathbb{M})$ is used as in previous work; we refer the reader to Section 2.3 of \citet{Raaijmakers20} and references therein for a more detailed description.

\section{EOS constraints}
\label{sec:eos}

In this Section we investigate the impact of the \citet{Riley21} mass-radius measurement for \joh on the dense matter EOS, both separately and when combined with previous constraints. 

\subsection{Radio mass measurement of PSR~J0740$+$6620}

Firstly, we constrain the EOS using the updated mass measurement of $2.08\pm 0.07$ \msol for PSR~J0740$+$6620 derived using radio timing \citep{Fonseca21}, and compare this to the constraints from the previously published mass of $2.14^{+0.1}_{-0.09}$ \msol \citep{Cromartie19}. In Figure \ref{fig:fig1} we show the posterior distribution on EOS parameters $\bm{\theta}$ when transformed to the mass-radius parameter space. We note that, as expected, the slightly lower updated mass measurement shifts the posterior distributions to lower maximum neutron star masses and lower radii, although the effect is almost negligible. Since the radio timing mass measurement is already incorporated in the joint mass-radius estimate from \NICER we will not use this measurement in the remainder of this work.

\subsection{GW170817 and GW190425}

The gravitational wave events GW170817 and GW190425 have so far been the only confirmed neutron star binary mergers during the recent observing runs of the LIGO/Virgo collaboration \citep{GWTC2}. Although both events have a non-negligible chance of being neutron star-black hole mergers (see, e.g., \citet{Yang:2017gfb, Ascenzi19, Coughlin19, Hinderer19} for GW170817 and e.g., \citet{Kyutoku20, Han20} for GW190425), in the following we will assume both objects to be neutron stars. We use the low-spin\footnote{The low-spin assumption is chosen to be consistent with measurements of spins in Galactic neutron star binaries that merge within a Hubble time.} posterior distributions on tidal deformability and mass ratio with the \texttt{IMRPhenomPv2\_NRTidal}\footnote{See Table 1 of \citet{GW170817sourceproperties} for a description of the waveform model.} waveform model \citep{Hannam:2013oca,Khan:2015jqa,Dietrich19a} for GW170817 and GW190425.
Furthermore we use the median chirp mass values of $\mathcal{M}_c=1.186$ \msol for GW170817\footnote{\url{https://dcc.ligo.org/LIGO-P1800370/public/}} and $\mathcal{M}_c=1.44$ \msol for GW190425\footnote{\url{https://dcc.ligo.org/LIGO-P2000223/public/}}.

\begin{figure}[t!]
\centering
\includegraphics[width=.47\textwidth]{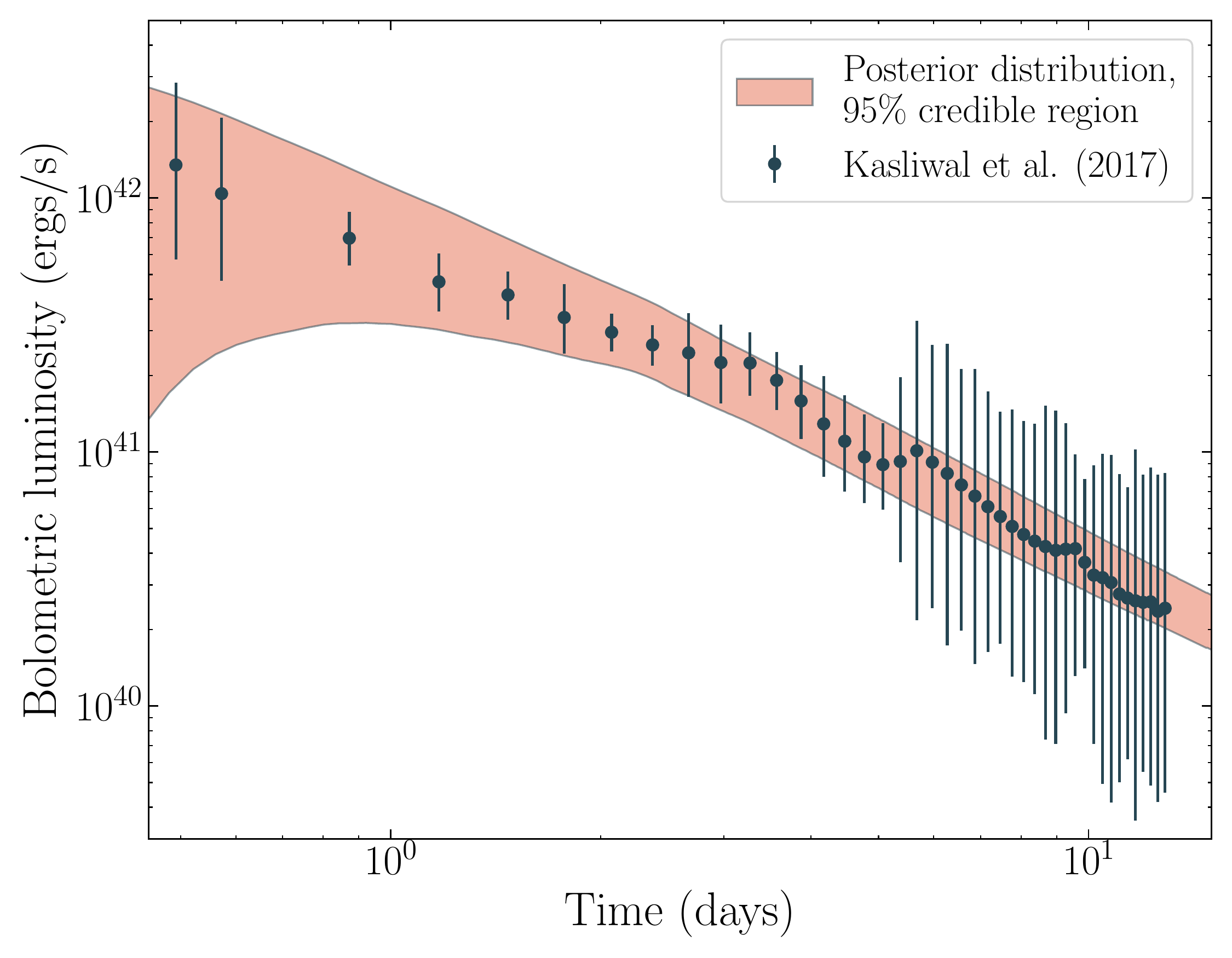}
\caption{In blue we show the bolometric luminosity of GW170817 from the data compiled in \citet{Kasliwal17}. The red band contains $95\%$ of the light curves of the posterior distribution when fitted with the model described in Section \ref{subsec:at2017gfo}.}
\label{fig:figEM}
\end{figure}

\begin{figure*}[t!]
\centering
\includegraphics[width=.95\textwidth]{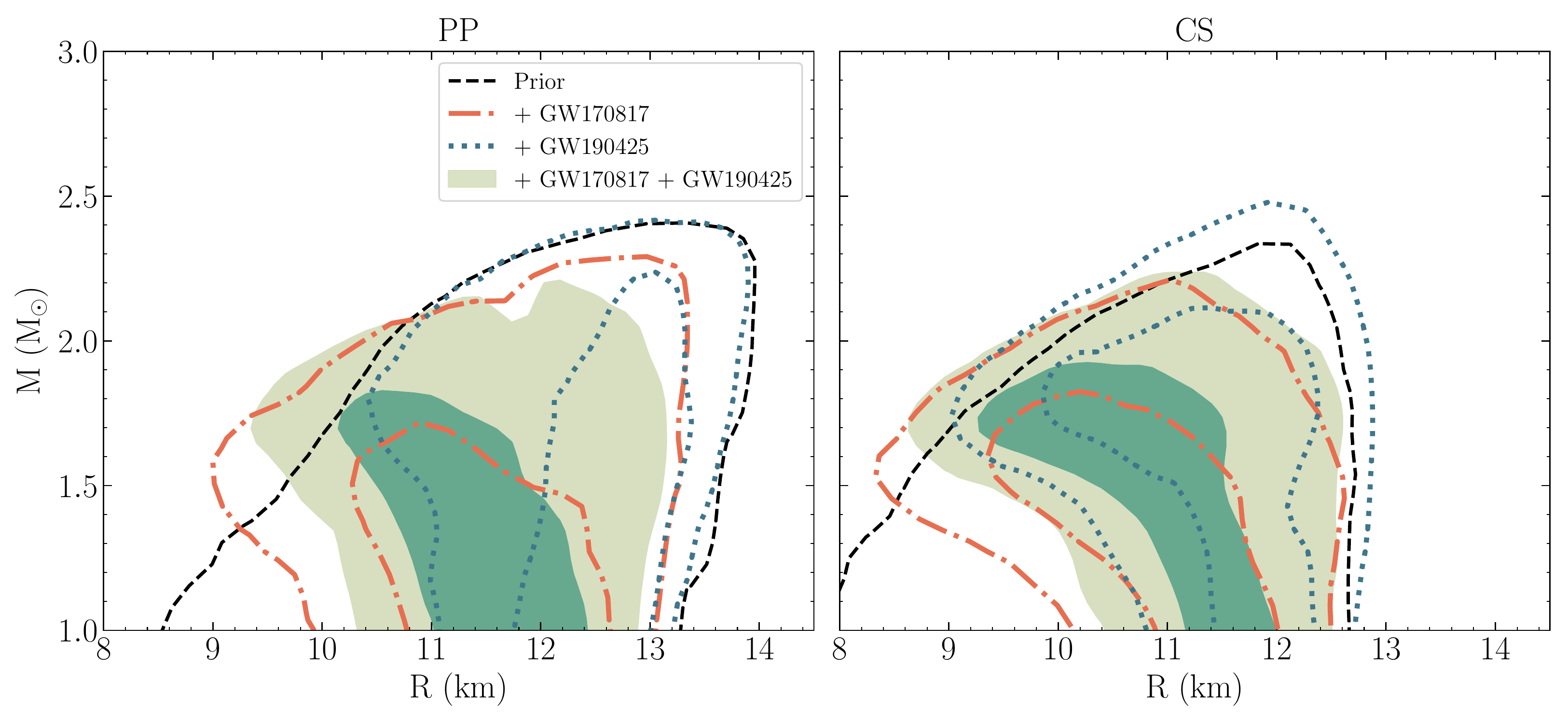}
\includegraphics[width=.95\textwidth]{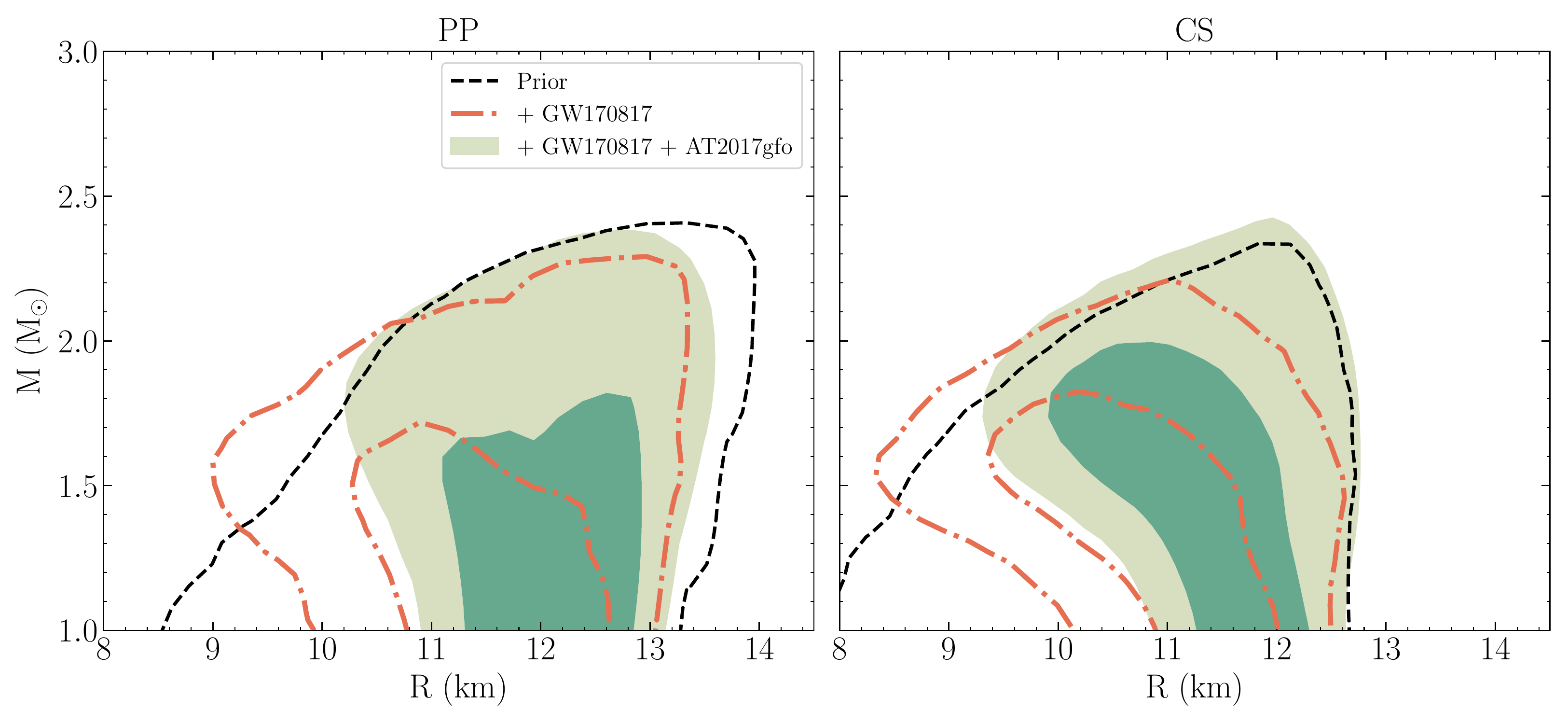}
\caption{\textit{Upper panels:} Constraints on the mass-radius relation of neutron stars, given the posterior distribution on EOS parameters $\bm{\theta}$ using the PP model (left) and CS model (right) when analyzing the gravitational wave events GW170817 \citep{GW170817discovery} and GW190425 \citep{GW190425}, both separately and combined. The estimated tidal deformability from GW170817 offers more posterior support for softer EOS, and thus lower radii. For GW190425 only weak upper limits could be set on the tidal deformability, but the relatively high estimated mass of the primary object disfavors softer EOS, as we are not considering any high-mass information from radio pulsars here. \textit{Lower panels:} The change in the posterior distribution on the EOS when including information from the kilonova associated with GW170817, AT2017gfo \citep{Kasliwal17}. The estimated mass that was ejected during the merger favors higher tidal deformabilities, and thus constrains the mass-radius space at low radii.}
\label{fig:fig2}
\end{figure*}

The upper panels of Figure \ref{fig:fig2} show the posterior distributions on the EOS for both events in the mass-radius space. We note that the constraints on tidal deformability from GW170817 give more support to softer EOS, although the $95\%$ credible region spans a relatively large range of radii. GW190425 only led to weak upper limits on the tidal deformability due to its low SNR and single-detector detection. The EOS is however constrained as a result of the high mass of the primary component (with $95\%$ credible range $1.60 - 1.87$ \msol), excluding EOS that do not support these masses.

\begin{figure*}[t!]
\centering
\includegraphics[width=.95\textwidth]{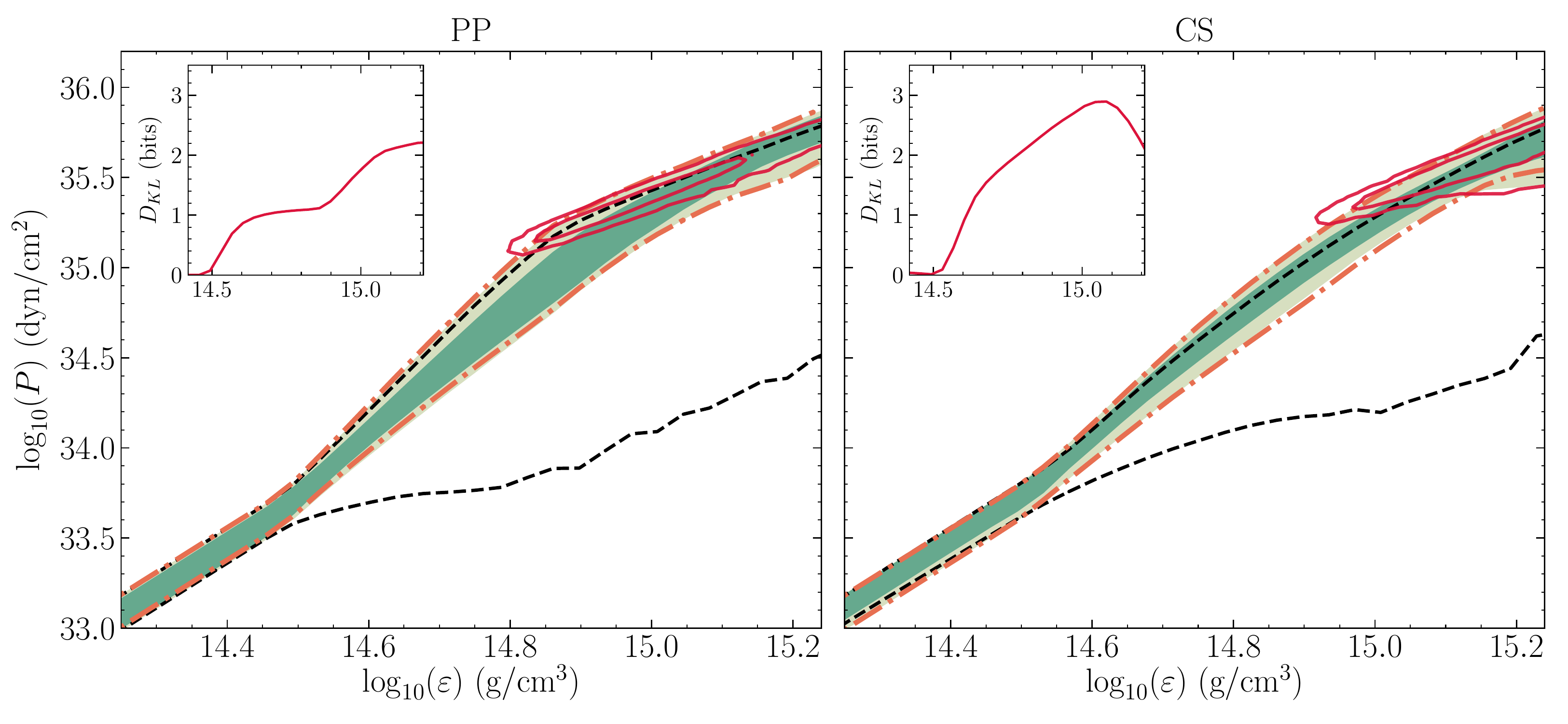}
\includegraphics[width=.95\textwidth]{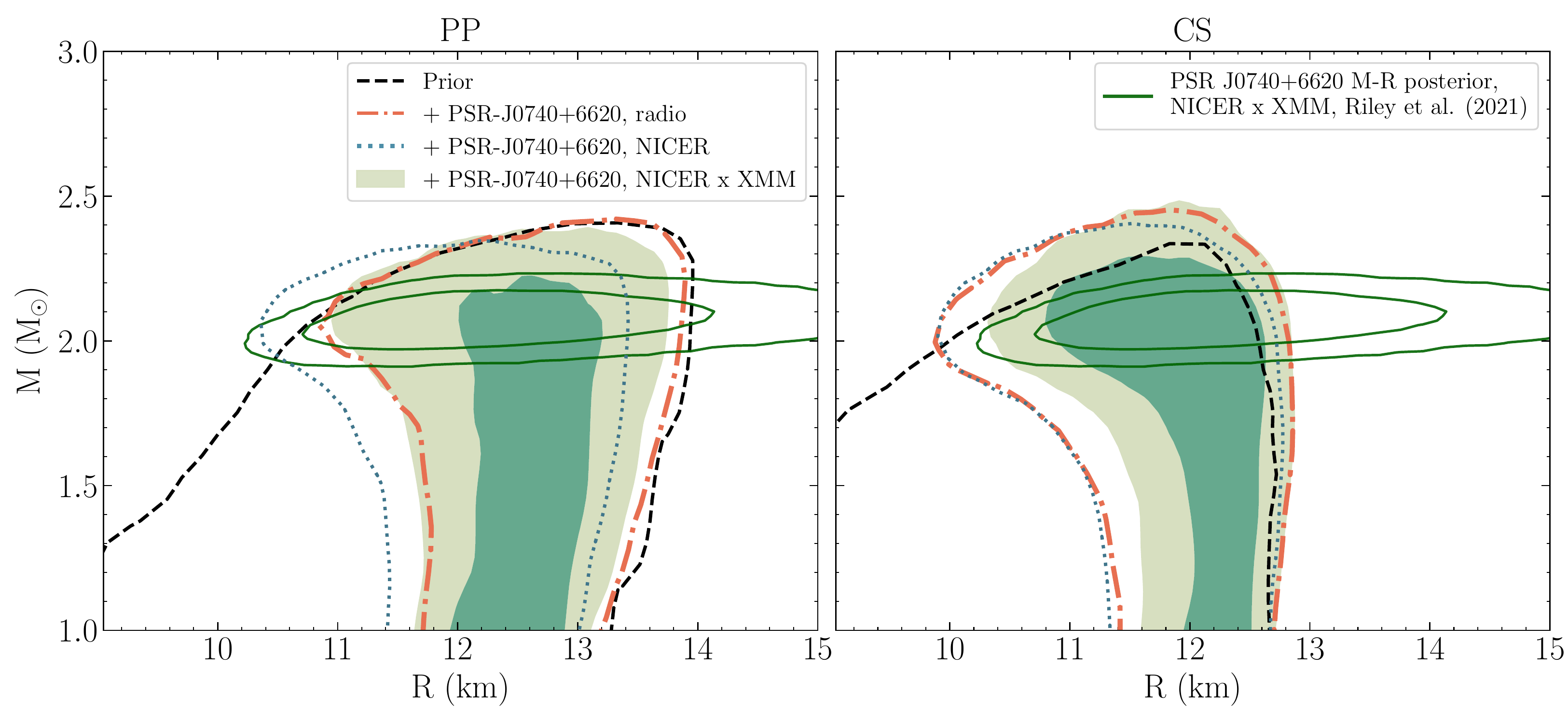}
\caption{\textit{Upper panels:} The $68\%$ and $95\%$ credible regions of the EOS given the mass-radius estimate of \joh by \citet{Riley21}, using the PP model (left) and CS model (right). The black dashed lines and orange dashed-dotted lines indicate the $95\%$ credible region of the prior and the constraints given the radio mass measurement of \joh by \citet{Fonseca21}, respectively. The red contour shows the posterior distribution on central energy density and pressure for this source, and in the inset we plot the KL-divergence as a function of energy density. \textit{Lower panels:} Same as upper panels but for the mass-radius space. Also shown in blue, dotted lines is the $95\%$ credible region of the EOS posterior distribution, when analyzing the result from \citet{Riley21} without the inclusion of the \textit{XMM}-dataset (so \NICER only). In addition, we show the mass-radius posterior for \joh by \citet{Riley21} as dark-green contours (68\% and 95\%). Note that when considering both \NICER and \textit{XMM} data, the posterior distribution (green shaded) is very close to the constraints obtained from the radio mass measurement of \joh (orange), due to this mass-radius posterior (dark green) showing support over an extended range of radii.}
\label{fig:fig6}
\end{figure*}

\begin{figure*}[t!]
\centering
\includegraphics[width=.95\textwidth]{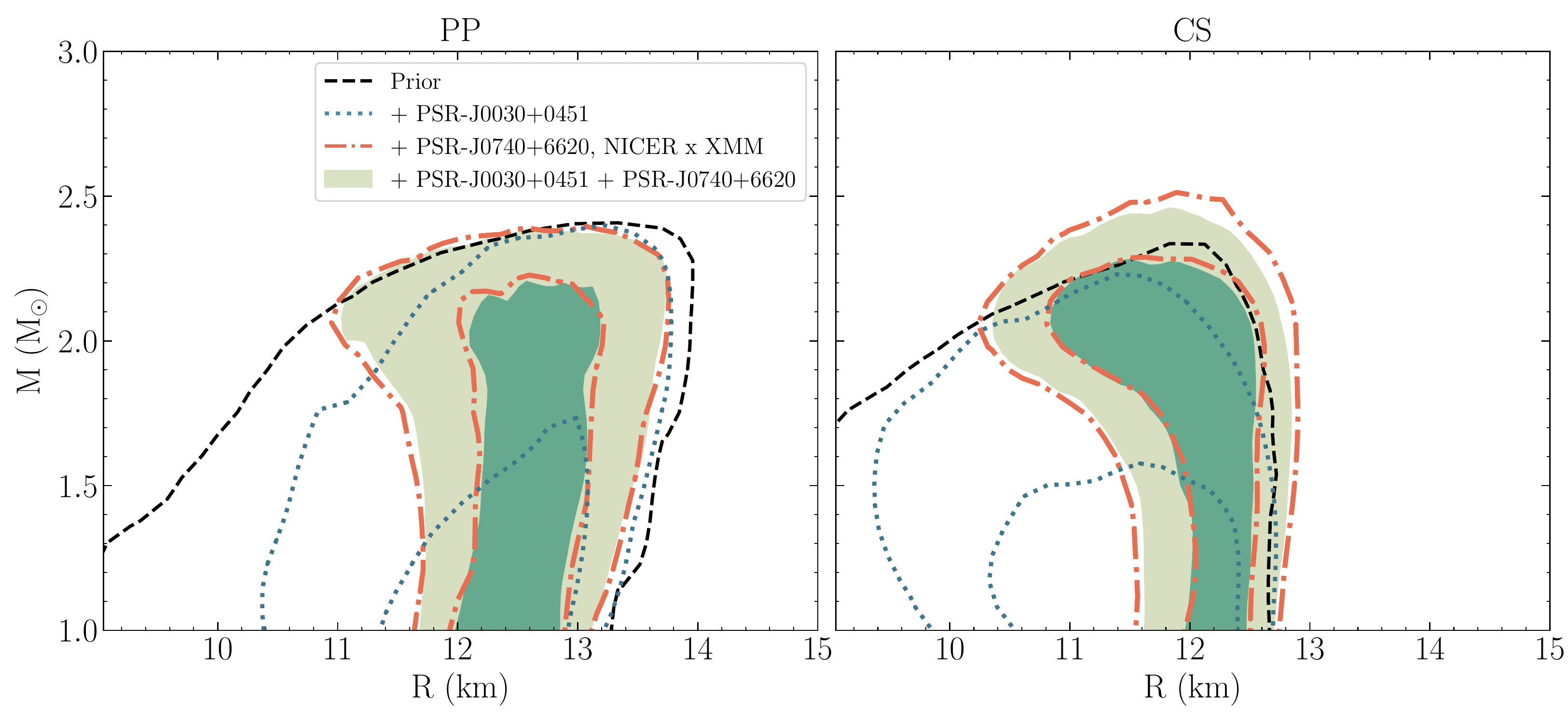}
\includegraphics[width=.95\textwidth]{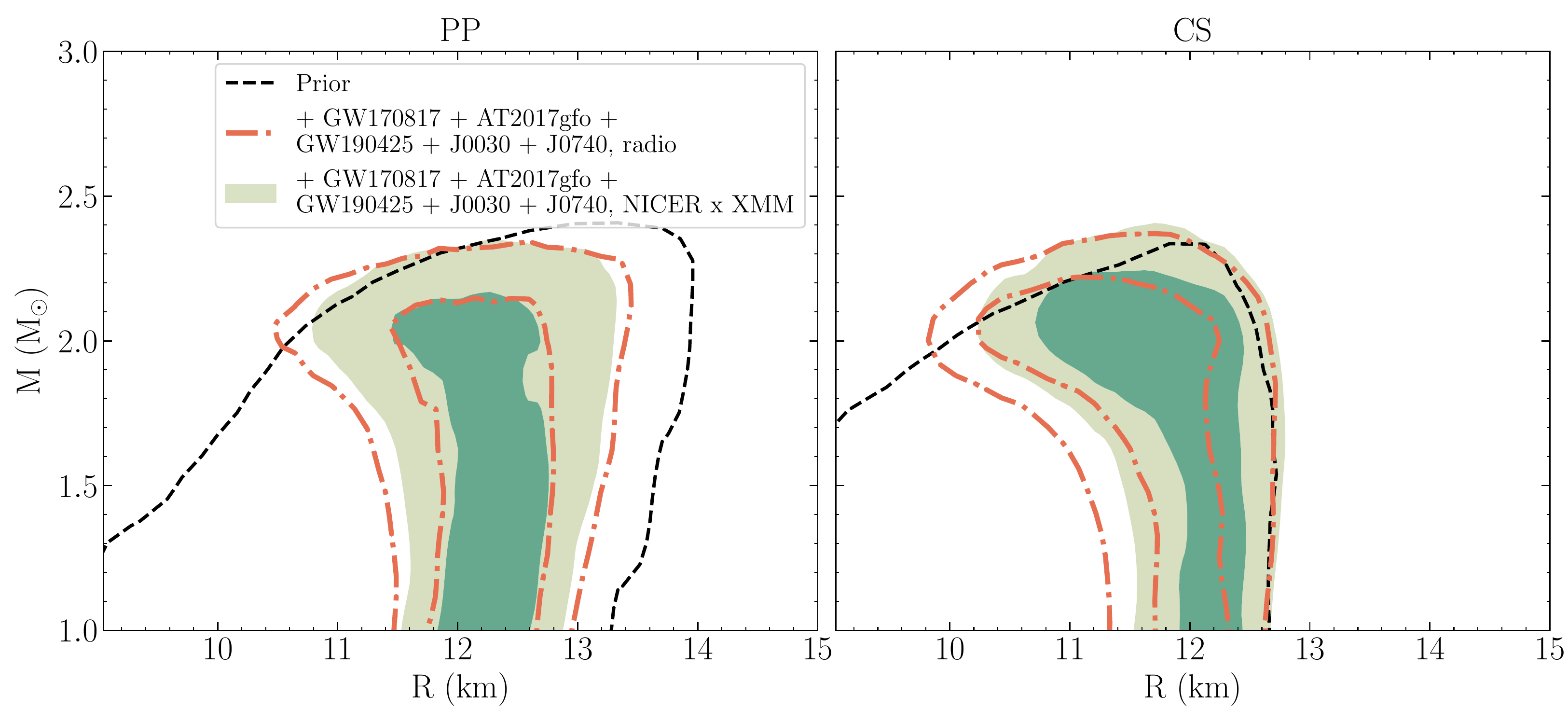}
\caption{\textit{Upper panels:} Constraints on the mass-radius space of neutron stars, given the posterior distribution of EOS parameters $\bm{\theta}$ using the PP model (left) and CS model (right). Shown are the $68\%$ and $95\%$ credible regions when analyzing \jdbl, \joh and the combination of the two pulsars. Note that the distribution of \jdbl is different than in \citet{Raaijmakers19}, because here we have not included any high-mass pulsar information. \textit{Lower panels:} Similar to upper panels, but when analyzing jointly mass-radius estimates from \joh \citep{Riley21}, \jdbl \citep{Riley19}, mass-tidal deformability estimates from GW170817 \citep{GW170817sourceproperties} and GW190425 \citep{GW190425} and the kilonova data of \citet{Kasliwal17} as described in Section \ref{subsec:at2017gfo}. Combined, we find the radius of a $1.4\,$\msol neutron star to be constrained to the 95\% credible ranges $12.33^{+0.76}_{-0.81}\,$km (PP model) and $12.18^{+0.56}_{-0.79}\,$km (CS model). To show the impact of the radius measurement of \joh we also plot the posterior distribution when analyzing combined constraints with only the $2.08$\msol mass measurement of \joh (orange dashed-dotted lines).}
\label{fig:fig6a}
\end{figure*}

\subsubsection{AT2017gfo}
\label{subsec:at2017gfo}

Following the detection of GW170817 an EM counterpart was observed across the frequency spectrum (see, e.g., \citet{LVCmultimessenger,GW170817grb} and references therein; \citet{ GW170817swope, Chornock17, Drout17, Hallinan17, Kasliwal17, Kasliwal19, Margutti17, Pian17, Smartt17, Troja17}). Of particular interest here is the thermal infrared-optical-ultraviolet transient powered by radioactive decay of r-process nucleosynthesis in the neutron-rich material ejected during merger; the so-called kilonova or macronova \citep[e.g.][]{Li98,Kulkarni05,2010MNRAS.406.2650M}. The kilonova properties depend on the mass, velocity, and composition of the ejected material, which in turn depend on the binary progenitor parameters such as the tidal deformability of the neutron stars. Using this connection it is possible to constrain the EOS from the kilonova light curve \citep[see, e.g.,][]{ Coughlin:2018miv,Radice:2018ozg, Hinderer19,Capano:2019eae,Dietrich20}. 

Here we analyze the bolometric luminosity of GW170817 \citep[as compiled in][]{Kasliwal17} via the new Bayesian framework outlined in \citet{Raaijmakers21}. We consider a two-component kilonova model, where the first component, the dynamical ejecta, is associated with material ejected through tidal forces and the shock interface between the two neutron stars \citep[see, e.g.,][and references therein]{Radice18a}. The second component is associated with neutrino-driven winds or material ejected through viscous forces. We connect the outflow properties of these components to the binary progenitor properties by using the formulae presented in \citet{Kruger20} for dynamical ejecta and disk mass, which are fitted to numerical simulations of compact mergers. The velocity of the dynamical ejecta is calculated using the formula in \citet{Coughlin19a}, while the velocity of the disk wind ejecta is left as a free parameter. The dynamical ejecta includes both material ejected through tidal forces and material ejected through shocks on the contact interface between the stars \citep[see, e.g.,][]{Sekiguchi16, Dietrich17, Tanaka20, Nedora21}. To distinguish these we consider two different opacities in the dynamical ejecta, corresponding to tidal tail and shock ejecta , where the latter is less neutron-rich compared to the tidal tail and thus has a lower opacity (see Table \ref{tab:tab1}). For simplicity we take a single opacity for the disk wind ejecta. The outflow properties are then connected to a bolometric luminosity through the semi-analytic light curve model by \citet{Hotokezaka19}. The priors on all parameters are shown in Table \ref{tab:tab1}.

The fit to the bolometric luminosity of AT2017gfo using the data compiled in \citet{Kasliwal17} is shown in Figure \ref{fig:figEM}, showing all datapoints to be contained within the $95\%$ credible region of the posterior distribution. In the lower panels of Figure \ref{fig:fig2} we show the updated prior distribution for the EOS with GW170817 and with the inclusion of AT2017gfo. The EM data gives more posterior support to stiffer over softer EOS, due to the estimated ejected mass requiring a neutron star with larger tidal deformability. The estimated radius of a $1.4$\,\msol neutron star for the PP and CS model is $12.12^{+1.10}_{-1.44}\,$km and $11.53^{+1.16}_{-1.15}\,$km, respectively, which is broadly consistent with multimessenger constraints obtained by other works \citep[see, e.g.,][]{Coughlin19a,Dietrich20, Capano:2019eae, Breschi21b, Nicholl21}. 
Important to note is that the EM modeling of the kilonova here is simplified and relies on a few assumptions that are known to affect results, such as spherical ejecta geometry \citep[see, e.g.,][]{Heinzel21, Korobkin21}, fixed nuclear heating rate \citep[see, e.g.,][]{Barnes20}, and an incomplete mapping between properties of the binary system and the ejecta outflows. It is also dependent on the choice of light curve modeling, where the distinction can be made between semi-analytic modeling \citep[such as in this work and, e.g.,][]{Breschi21b, Nicholl21} and interpolating between radiative transfer simulations \citep[e.g.,][]{Coughlin:2018miv, Dietrich20}. 
We use a semi-analytical model from \citet{Hotokezaka19}, which for the current statistical uncertainty in gravitational wave parameter estimation and uncertainty in light curve observations produces consistent results to full radiative transport models \citep[see, e.g.,][]{Coughlin_2019, Coughlin_2020}, although this will change in the future with improved gravitational wave detectors and optical telescopes.

\begin{table}[h!]
\caption{The parameters used in the model described in Section \ref{subsec:at2017gfo} and their prior support in the analysis of AT2017gfo. The notation $U(a,b)$ here means uniformly drawn between boundaries $a$ and $b$.}
\centering
\begin{tabular}{@{}lll@{}}
\toprule
Parameters                                   & Prior density and support     \\ \midrule
\multicolumn{2}{c}{\textit{Binary properties}}                                                                                                     \\ \midrule
$\mathcal{M}_c$ [M$_{\odot}$]                       & $\sim$ U(1.18, 1.2)           \\
q                      & $\sim$ U(0.2, 1)                       \\
$\Lambda_{1}$                               & $\sim$ U(0, 2500)               \\
$\Lambda_{2}$                               & $\sim$ U(0, 2500)                \\ \midrule
\multicolumn{2}{c}{\textit{Ejecta and light curve properties}}                                                                                     \\ \midrule
M$_{\rm{dyn}}$ [M$_{\odot}$]                &  Eq. (2) \citet{Raaijmakers21}  \\
v$_{\rm{dyn}}$ [c]                          &          Eq. (D5) \citet{Coughlin19a}           \\
$v_{\rm{min, dyn}}$ [c]                     & $\sim$ U(0.1, 1.0) v$_{\rm{dyn}}$              \\
$v_{\rm{max, dyn}}$ [c]                     & $\sim$ U(1.5, 2.5) v$_{\rm{dyn}}$                \\
v$_{\kappa}$ [c]                           & $\sim$ U(v$_{\rm{min, dyn}}$, v$_{\rm{max, dyn}}$) \\
$\kappa_{\rm{low}}$ [cm$^2$ g$^{-1}$]      & $\sim$ U(0.1, 5)                                \\
$\kappa_{\rm{high}}$ [cm$^2$ g$^{-1}$]     & $\sim$ U(5, 30)                                \\ \midrule
M$_{\rm{wind}}$ [M$_{\odot}$]               &  Eq. (10) \citet{Raaijmakers21}               \\
v$_{\rm{wind}}$ [c]                         & $\sim$ U(0.03, 0.15)                                 \\
$v_{\rm{min, wind}}$ [c]                   & $\sim$ U(0.1, 1.0) v$_{\rm{wind}}$                 \\
$v_{\rm{max, wind}}$ [c]                   & $\sim$ U(1.5, 2.0) v$_{\rm{wind}}$    \\
$\kappa_{\rm{wind}}$ [cm$^2$ g$^{-1}$]       & $\sim$ U(0.1, 5)                \\ \bottomrule 
\end{tabular}
\label{tab:tab1}
\end{table}

\subsection{\NICER mass-radius and multimessenger constraints}

Next we study the constraints on the EOS from the new mass-radius estimate of \joh using data from \NICER and \textit{XMM}, presented in \citet{Riley21}. They find a radius of $12.39_{-0.98}^{+1.30}$\,km, and a mass of $2.072_{-0.066}^{+0.067}$\,M$_{\odot}$, where the upper and lower limit bound the $68\%$ credible regions. The EOS results are shown in Figure \ref{fig:fig6}, both in energy density-pressure and mass-radius space. From the Kullback-Leibler (KL)-divergence \citep{Kullback1951} plotted as a function of energy density in the upper insets, we find that especially at higher energy densities there is a significant information gain from prior-to-posterior. Note that similar but, especially for the CS model, broader constraints are found for the posterior distribution when only using the radio mass measurement of \joh, as indicated by the orange dashed-dotted lines. This is a result of the mass-radius estimate of \joh being very consistent with our prior ranges informed by low-density chiral EFT calculations. The chiral EFT calculations do exclude however stiffer EOS with radii $>14$\,km, where the mass-radius posterior of \joh has non-negligible posterior support. For the CS model this effect is stronger as additional constraints on the speed of sound at $1.5n_0$ in the CS model lead to overall smaller radii than in the PP model (see Section 2.3 of \citealt{Raaijmakers20}). 

In the mass-radius space we also plot the EOS constraints given the joint \NICER mass-radius estimate excluding the \textit{XMM} data. For this analysis \citet{Riley21} report a value of $11.29_{-0.81}^{+1.20}$\,km for the radius, and $2.078_{-0.063}^{+0.066}$\,M$_{\odot}$ for the mass. As this joint mass-radius estimate has slightly more posterior support for lower radii, the corresponding EOS constraints suggest a softening of the EOS at high densities. These results should be interpreted with caution however, because the \NICER-only analysis leads to an under-prediction of the background (the contribution from instrumental or astrophysical background to the unpulsed component of the pulse profile). This results in more of the unpulsed component being attributed to the hot regions via high compactness solutions.  The \textit{XMM} data show that a larger component of the unpulsed emission must come from true background, eliminating these high compactness solutions and increasing the inferred radius in the joint \NICER-\textit{XMM} analysis (see also Section 4.2 in \citet{Riley21}).  

Finally, in Figure \ref{fig:fig6a} we show the constraints on the EOS from \joh, \jdbl (first derived in \citealt{Raaijmakers19}, but here no information on high-mass pulsars is included) and the combination of the two pulsars. Note that for the combined constraints, most of the information comes from \joh, since the $68\%$ credible region of the mass-radius posterior of \jdbl covers a broad range in radii that are consistent with the EOS constraints from \joh.

In the lower panels of Figure \ref{fig:fig6a} we show the combined constraints on the EOS including mass-radius estimates from \joh \citep{Riley21}, \jdbl \citep{Riley19} and mass-tidal deformability estimates from GW170817 \citep{GW170817sourceproperties} and GW190425 \citep{GW190425} and the kilonova AT2017gfo \citep{Kasliwal17}. We find that especially the pulsar mass-radius estimates by \NICER favor stiffer EOS, as well as GW170817 when the associated kilonova AT2017gfo \citep{Kasliwal17} is included. The weak constraints from GW190425 on the tidal deformability are also broadly consistent with the constraints coming from the other sources. As a comparison we show the posterior distribution when combining all analyses excluding the mass-radius estimate of \joh, but with the radio mass measurement of \citet{Fonseca21}. We note that the additional radius information on \joh constrains the softer EOS, especially for the CS model.

\begin{figure}[t!]
\centering
\includegraphics[width=.48\textwidth]{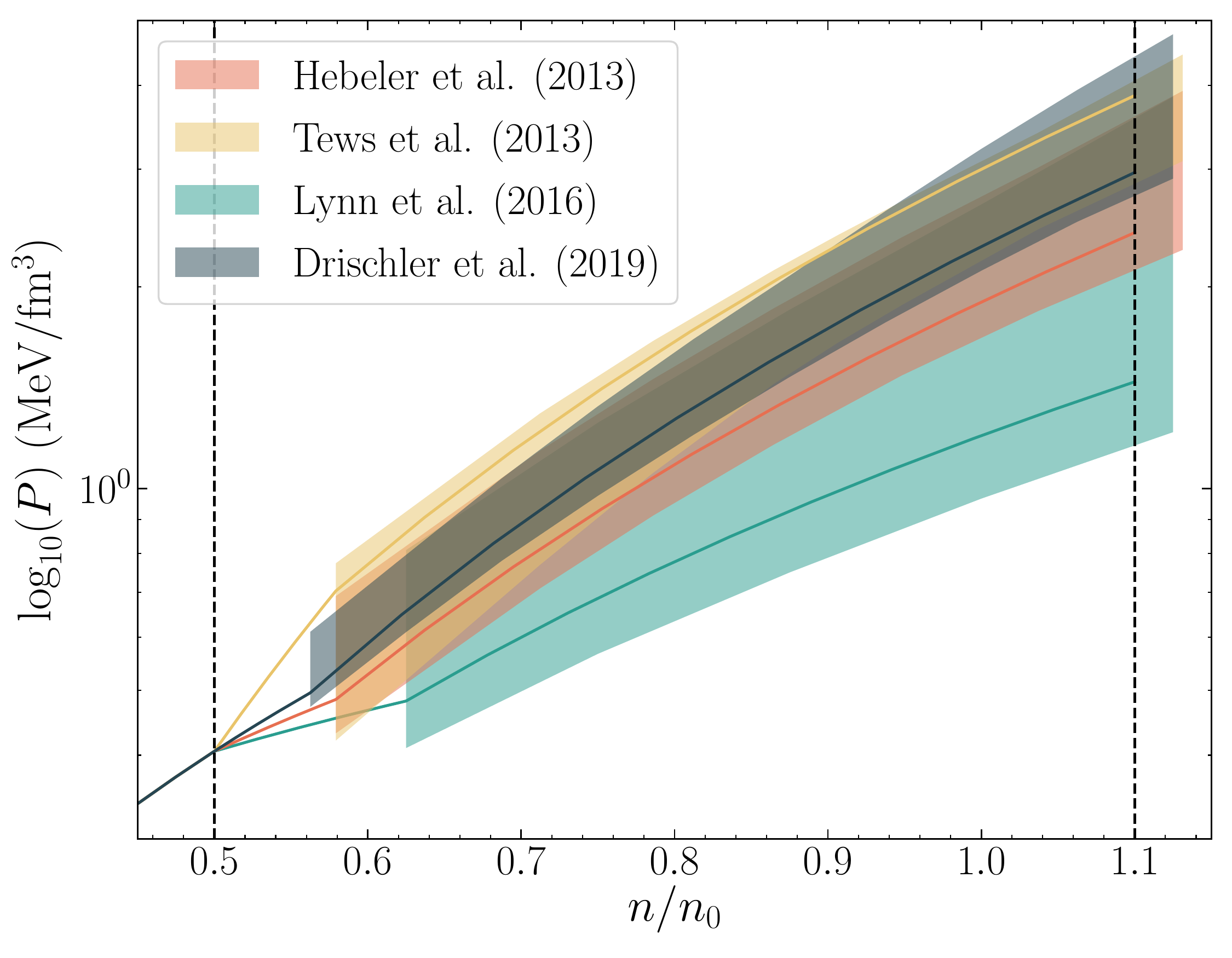}
\caption{Different chiral EFT bands for the pressure of neutron star matter at nuclear densities, $n/n_0$ in units of saturation density $n_0 = 0.16$\,fm$^{-3}$, and their matching to the BPS crust EOS at $0.5 n_0$. The different bands are based on microscopic calculations of neutron matter from \citet{Hebeler13}, \citet{Tews13}, \citet{Lynn16} and \citet{Drischler19} and include beta equilibrium (with protons and electrons) following the construction in \citet{Hebeler13}. The four chiral EFT calculations are considered between $0.5 n_0$ and $1.1 n_0$ in the analyses presented in Section \ref{sec:ceft}. Also shown are examples of the fit we use to approximate the EOS within these uncertainty bands, see Eq.~(\ref{eq:ceft}), and connect to the BPS crust EOS. For a comparison of the chiral EFT bands in pure neutron matter, see Figure~1 in \citet{Huth21}.}
\label{fig:fig5}
\end{figure}

\begin{figure*}[t!]
\centering
\includegraphics[width=.95\textwidth]{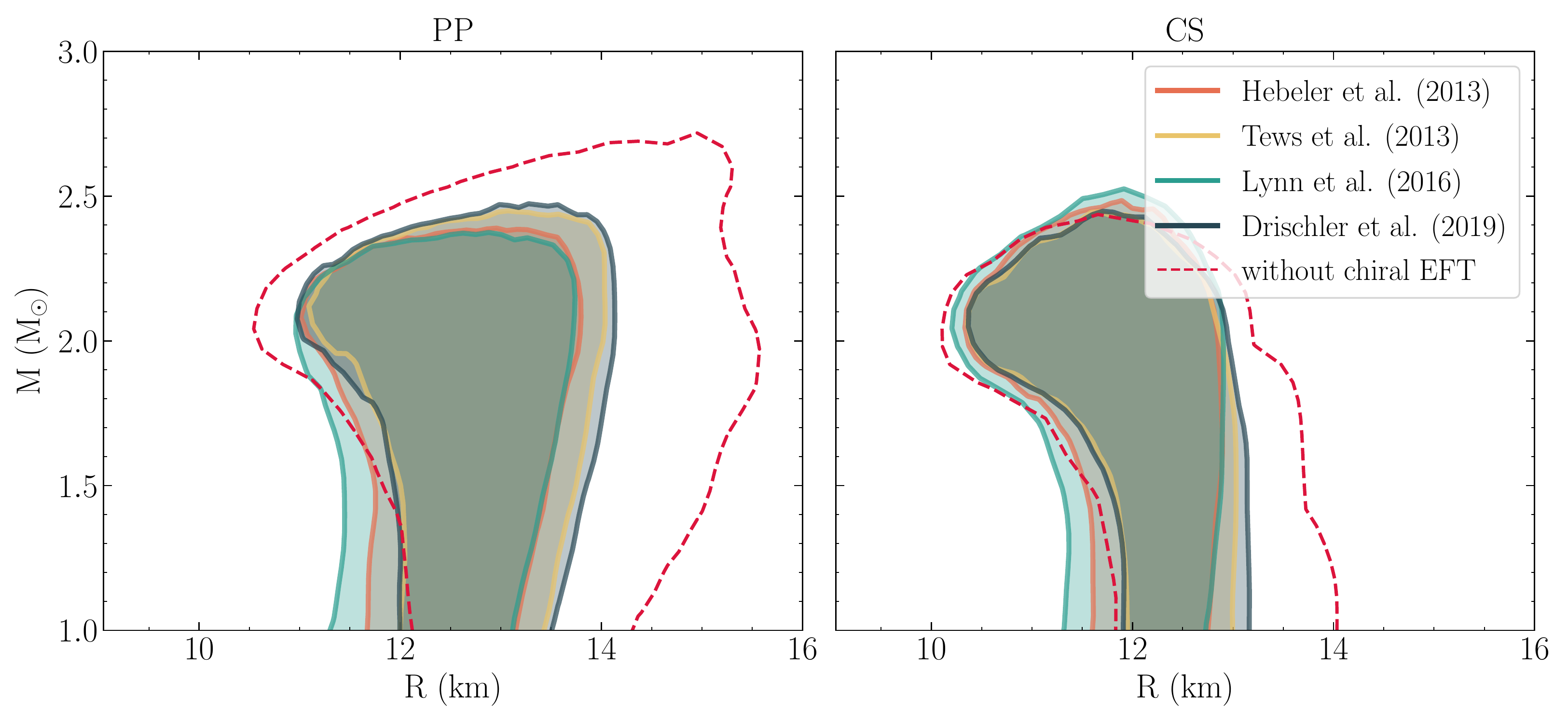}
\includegraphics[width=.95\textwidth]{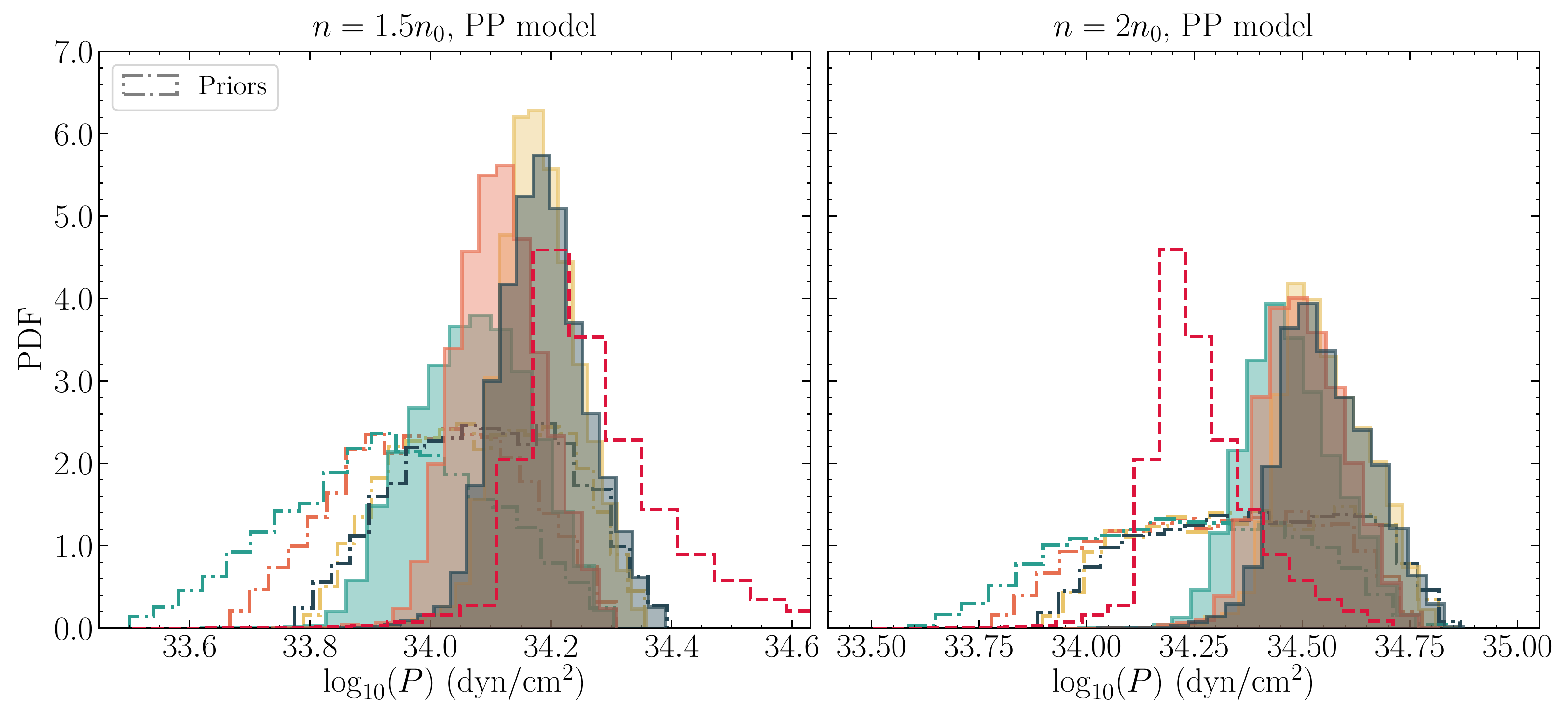}
\caption{\textit{Upper panels:} $95\%$ credible region for the mass-radius space given the mass-radius estimate of \joh by \citet{Riley21}, using the PP model (left) and CS model (right). The different results correspond to using the four different chiral EFT calculations between $0.5$ and $1.1 n_0$ as shown in Fig.~\ref{fig:fig5}. Moroever, the red, dashed lines correspond to the $95\%$ credible region, if the PP or CS parameterization is used down to $0.5 n_0$, i.e., immediately following the BPS crust, so that no information from chiral EFT is used. \textit{Lower panels:} Marginalized posterior distributions for the pressure $P$ above saturation density, at density $n = 1.5 n_0$ (left) and $n = 2 n_0$ (right) above the chiral EFT bands.}
\label{fig:fig4}
\end{figure*}

\section{Sensitivity of posteriors to nuclear constraints at low densities}
\label{sec:ceft}

To investigate the impact of the EOS constraints from nuclear physics we compare our analysis of \joh using four different chiral EFT uncertainty bands. All bands are based on microscopic calculations for pure neutron matter, which are then extended to neutron star matter in beta-equilibrium using the formalism discussed in \citet{Hebeler13}.
In order to improve the description of all employed EOSs, we generalized the density dependence of the energy-density functional [see Eq.~(2) in \citet{Hebeler13}] by enlarging the range of the exponent $\gamma$ to $\gamma \in \left[ 1.2, 2.5 \right]$.

The results from \citet{Hebeler13} formed the basis of our previous studies \citep{Raaijmakers19,Raaijmakers20}. The calculations for pure neutron matter were initially performed in \citet{Hebeler10a} using many-body perturbation theory, while the uncertainty band results mainly from variations of the couplings involved in three-nucleon interactions. Second, in \citet{Tews13} the calculations for neutron matter were improved by including for the first time all \text{two-,} three-, and four-neutron interactions to next-to-next-to-next-to-leading order (N$^3$LO), which are predicted in a parameter-free way for neutron matter (see, e.g., \citealt{Hebe15ARNPS,Hebeler:2020ocj} for reviews). Third, in \citet{Drischler19} the calculations were further optimized by improving the treatment of three-nucleon interactions and extending the many-body expansion to higher orders. In addition, the EOS uncertainty bands also include effects from variations of regulator scales in state-of-the-art nucleon-nucleon and three-nucleon interactions. In this work, we use the combined 450\,MeV and 500\,MeV N$^3$LO uncertainty bands from \citet{Drischler19}. Finally, we include results of \citet{Lynn16}. These were obtained by nonperturbative quantum Monte-Carlo simulations of neutron matter at next-to-next-to-leading order (N$^2$LO). This represents a completely different many-body method than those used for the other three bands, and the results of \citet{Lynn16} are also based on a different set of local two- and three-nucleon interactions derived from chiral EFT.

\begin{table*}[t!]
\caption{Key quantities from the posterior distributions obtained in Sections \ref{sec:eos} and \ref{sec:ceft}: The radius of a $1.4$\,\msol, $1.6$\,\msol, and $1.8$\,\msol  neutron star, as well as $\Delta R = R_{2} - R_{1.4}$, and the maximum mass of a non-rotating neutron star M$_{\rm TOV}$. For the analyses of Section \ref{sec:ceft}, we also show the inferred central energy density $\varepsilon_c$, the corresponding central pressure P$_{c}$, and the Bayes' factor $K$ comparing with the model using the chiral EFT band from \citet{Hebeler13}. The first four column results are for the different chiral EFT bands from \citet{Hebeler13} (Heb 13), \citet{Tews13} (Tews 13), \citet{Lynn16} (Lynn 16), and \citet{Drischler19} (Dri 19), while all other results are for the baseline inference using Heb 13. The column ``Combined with'' refers to the \NICER $\times$ \textit{XMM} analysis of \joh, the \NICER analysis of \jdbl and multimessenger constraints combined, while in the column ``Combined without'' the \NICER $\times$ \textit{XMM} analysis of \joh is replaced with just the radio mass measurement by \citet{Fonseca21}. The radii are given in km, M$_{\rm TOV}$ in \msol, and $\varepsilon_{c}$ and P$_c$ in g/cm$^{3}$ and dyn/cm$^{2}$, respectively. The upper and lower values correspond to the $95\%$ credible interval.}
\centering
\begin{tabular}{@{}cccccccccc@{}}
\toprule
 & \multicolumn{4}{c}{\joh, \NICER x XMM}  & PSR J0030 & GW170817 & GW170817  & Combined & Combined\\ 
  & Heb 13 & Tews 13 & Lynn 16  & Dri 19 & +0451 & + GW190425 & + AT2017gfo & without & with \\ \midrule
  \multicolumn{9}{c}{\textit{PP model}} \\ \midrule
 $R_{1.4}$ & $12.56^{+0.80}_{-0.91}$ & $12.85^{+0.77}_{-0.95}$ & $12.35^{+0.96}_{-0.98}$ & $12.87^{+0.85}_{-0.98}$ & $12.35^{+0.99}_{-1.99}$ & $11.51^{+1.51}_{-1.47}$ & $12.12^{+1.10}_{-1.44}$ & $12.30^{+0.72}_{-0.76}$ & $12.33^{+0.76}_{-0.81}$ \\
 R$_{1.6}$ & $12.60^{+0.87}_{-1.00}$ & $12.87^{+0.87}_{-1.05}$ & $12.40^{+1.03}_{-1.04}$ & $12.90^{+0.94}_{-1.08}$ & $12.50^{+0.96}_{-2.08}$& $11.43^{+1.68}_{-1.53}$ & $12.10^{+1.23}_{-1.69}$ & $12.32^{+0.92}_{-0.99}$ & $12.35^{+0.83}_{-0.90}$ \\
 R$_{1.8}$ & $12.62^{+0.98}_{-1.19}$ & $12.86^{+1.00}_{-1.26}$ & $12.42^{+1.13}_{-1.19}$ & $12.89^{+1.08}_{-1.29}$ & $12.68^{+0.94}_{-1.99}$ & $11.65^{+1.64}_{-1.80}$ & $12.22^{+1.26}_{-1.91}$ & $12.29^{+1.06}_{-1.19}$ & $12.33^{+0.97}_{-1.06}$ \\
  $\Delta R$ & $-0.24^{+0.65}_{-1.04}$ & $-0.17^{+1.26}_{-1.21}$ & $-0.22^{+0.60}_{-1.05}$ & $-0.45^{+1.14}_{-1.07}$ & $-0.13^{+0.76}_{-1.02}$& $-0.35^{+0.80}_{-1.09}$ & $-0.26^{+0.77}_{-1.14}$ & $-0.30^{+0.64}_{-1.06}$ & $-0.29^{+0.61}_{-0.98}$ \\
 M$_{\rm TOV}$ & $2.26^{+0.15}_{-0.23}$ & $2.33^{+0.14}_{-0.30}$ & $2.22^{+0.19}_{-0.21}$ & $2.33^{+0.18}_{-0.31}$ & $1.74^{+0.66}_{-0.57}$& $1.84^{+0.51}_{-0.17}$ & $1.96^{+0.42}_{-0.44}$ &$2.23^{+0.15}_{-0.23}$ & $2.23^{+0.14}_{-0.23}$ \\ 
 $\log_{10}(\varepsilon_{c})$ & $14.99^{+0.27}_{-0.16}$ & $14.99^{+0.28}_{-0.18}$ & $15.00^{+0.26}_{-0.16}$ & $14.99^{+0.28}_{-0.19}$ &$14.86^{+0.28}_{-0.13}$ & -  & -  & - \\
 $\log_{10}$(P$_{c})$ & $35.39^{+0.39}_{-0.24}$ & $35.37^{+0.41}_{-0.26}$ &$35.41^{+0.37}_{-0.25}$  & $35.37^{+0.43}_{-0.28}$ & $34.92^{+0.30}_{-0.21}$& - & - &  -\\ 
  K & 1.00 & 0.89 & 1.00 & 0.85 & -& - & -& -  \\ \midrule
  \multicolumn{9}{c}{\textit{CS model}} \\ \midrule
R$_{1.4}$ & $12.27^{+0.54}_{-0.90}$ & $12.49^{+0.49}_{-0.87}$ & $12.16^{+0.63}_{-0.97}$ & $12.56^{+0.51}_{-0.92}$ &$11.51^{+1.12}_{-1.90}$ & $11.18^{+1.33}_{-1.51}$ & $11.53^{+1.16}_{-1.15}$ & $11.98^{+0.63}_{-0.71}$ & $12.18^{+0.56}_{-0.79}$ \\
 R$_{1.6}$ & $12.25^{+0.59}_{-0.94}$ & $12.43^{+0.55}_{-0.92}$ & $12.16^{+0.66}_{-0.99}$ & $12.50^{+0.55}_{-0.96}$ &$11.48^{+1.20}_{-1.92}$ & $10.92^{+1.58}_{-1.56}$ & $11.33^{+1.38}_{-1.41}$ & $11.91^{+0.78}_{-0.94}$ & $12.14^{+0.61}_{-0.84}$ \\
 R$_{1.8}$ & $12.14^{+0.69}_{-1.05}$ & $12.27^{+0.66}_{-1.04}$ & $12.08^{+0.74}_{-1.07}$ & $12.33^{+0.65}_{-1.06}$ & $11.52^{+1.23}_{-1.77}$& $10.85^{+1.73}_{-1.47}$ & $11.34^{+1.44}_{-1.65}$ & $11.72^{+0.96}_{-1.06}$ & $12.00^{+0.74}_{-0.96}$ \\
 $\Delta R$ & $-0.69^{+1.10}_{-1.02}$ & $-0.72^{+1.12}_{-1.08}$ & $-0.58^{+1.03}_{-1.08}$ & $-1.06^{+1.46}_{-0.83}$ &$-0.93^{+1.31}_{-0.86}$ & $-0.93^{+1.36}_{-0.83}$ & $-0.81^{+1.22}_{-0.92}$ & $-0.91^{+1.15}_{-0.85}$ & $-0.74^{+1.09}_{-0.95}$ \\
 M$_{\rm TOV}$  & $2.13^{+0.33}_{-0.16}$ & $2.13^{+0.29}_{-0.18}$ & $2.14^{+0.34}_{-0.17}$ & $2.12^{+0.31}_{-0.16}$ &$1.46^{+0.82}_{-0.42}$ & $1.81^{+0.45}_{-0.15}$ & $1.85^{+0.56}_{-0.30}$ & $2.09^{+0.26}_{-0.15}$ & $2.11^{+0.29}_{-0.16}$ \\ 
 $\log_{10}(\varepsilon_{c})$  & $15.19^{+0.21}_{-0.20}$ & $15.19^{+0.20}_{-0.20}$ & $15.18^{+0.21}_{-0.21}$ & $15.20^{+0.19}_{-0.20}$ &$15.03^{+0.33}_{-0.21}$ & -  & -  & - \\
 $\log_{10}$(P$_{c})$ & $35.61^{+0.30}_{-0.27}$ & $35.62^{+0.30}_{-0.28}$ &$35.60^{+0.30}_{-0.28}$  & $35.63^{+0.30}_{-0.29}$ & $35.05^{+0.30}_{-0.24}$& - & - &  -\\ 
  K & 1.00 & 1.04 & 0.92 & 1.05 & - & - & -& -\\\bottomrule
\end{tabular}
\label{tab:tab2}
\end{table*}

Similar to \citet{Raaijmakers20} we approximate the EOS within these bands with a single polytrope $P = N n^\Gamma$. However, to obtain a better fit to the additional bands considered here, we vary the polytropic index $\Gamma$ as a function of the normalization $N$, 
\begin{equation}
\label{eq:ceft}
    \Gamma (N) = \frac{(N - N_{\rm min})}{(N_{\rm max} - N_{\rm min})}  (\Gamma_{\rm max} - \Gamma_{\rm min}) + \Gamma_{\rm min}\,,
\end{equation}
where $N_{\rm min/max}$ and $\Gamma_{\rm min/max}$ are determined by fitting a polytrope to the lower and upper bound of the band. In Figure \ref{fig:fig5} we show the four different bands for the pressure of neutron star matter with an example of the fit through each band. This shows the consistency of these different chiral EFT calculations, with different methods, interactions, and approximations. The first point of the band where $n/n_0 > 0.5$ is matched to the BPS crust EOS at $0.5 n_0$ via a linear interpolation.

We study the dependence of the EOS constraints on the different chiral EFT bands by inferring the EOS from the mass-radius estimate of \joh using each band and both high-density parameterizations. The results are shown in Figure \ref{fig:fig4}. We also show the $95\%$ credible region of the updated prior distribution when directly joining the PP or CS high-density parameterization to the crust EOS at $0.5\rho_{ns}$. As expected the chiral EFT calculations mostly exclude stiffer EOS. While the different chiral EFT bands yield very good agreement on the upper bound of the radius estimates, the lower bound on the radius does slightly depend on the chiral EFT band used, especially at lower neutron star masses, depending on how soft the chiral EFT band is (see Figure~\ref{fig:fig5}). 

In the lower panels of Figure \ref{fig:fig4} we also show the posterior distributions on the pressure at densities $n = 1.5 n_0$ and $n = 2 n_0$ above the chiral EFT bands. These results demonstrate that the \joh mass-radius measurement systematically prefers higher pressures at these densities compared to the corresponding prior distributions of each chiral EFT band. Furthermore, the posteriors at $n = 2 n_0$ agree very well for all chiral EFT bands and are peaked around $P \sim 10^{34.5} \text{dyn}/\text{cm}^2 \sim 20 \, \text{MeV}/\text{fm}^3$.

\section{Discussion}

In this Letter, we have investigated the constraints on the EOS posed by the new joint mass-radius estimate from \NICER $\times$ \textit{XMM} data \citep{Riley21}, and compared and combined with multimessenger EOS constraints from radio timing, gravitational wave mergers and their counterparts, and the previous \jdbl mass-radius estimate by \NICER. In Table \ref{tab:tab2} we summarize the results obtained in Sections \ref{sec:eos} and \ref{sec:ceft} for the constraints on the radius of a $1.4, 1.6$ and $1.8$\,\msol neutron star, as well as $\Delta R = R_{2} - R_{1.4}$, and the maximum mass of a non-rotating neutron star M$_{\rm TOV}$, as well as the constraints on the central energy density and pressure for \joh. 

\subsection{Implications for nuclear physics}

We have studied the sensitivity of the EOS constraints from \joh using four different low-density EOS calculations from chiral EFT (see Section \ref{sec:ceft}). From the results presented in Figure \ref{fig:fig4} and Table \ref{tab:tab2} we conclude that the constraints on the EOS are only weakly dependent on the choice of low-density calculations, although small differences exist at lower radii. Assuming all four low-density calculations to be equally probable, we can compute the Bayes' factor $K$ by taking the ratio of the evidence of each \MultiNest run, and assess whether one model is preferred over another by the data of \joh. We list the Bayes' factors in Table \ref{tab:tab2}, where each model is compared to using the chiral EFT band from \citet{Hebeler13}. All values are close to one, indicating that there is no substantial support for one model over the other, based on the mass-radius estimate of \joh. These results are consistent with the observation that predictions for pure neutron matter are well constrained by modern nuclear forces derived within chiral EFT \citep{Huth21,Hebeler:2020ocj}.

Also shown in Table \ref{tab:tab2} are the values of $\Delta R = R_{2} - R_{1.4}$, the difference in radius of a $2$\,\msol and $1.4$\,\msol neutron star. As pointed out by \citet{Drischler20a}, the value of $\Delta R$, if positive, can give an indication that possibly unusual stiffening happens at high densities. We find however all values to be consistent with the mean $\Delta R$ being negative, but due to the broad uncertainty no conclusive statements can be made.

\subsection{Implications for maximum mass} 

\begin{figure*}[t!]
\centering
\includegraphics[width=.95\textwidth]{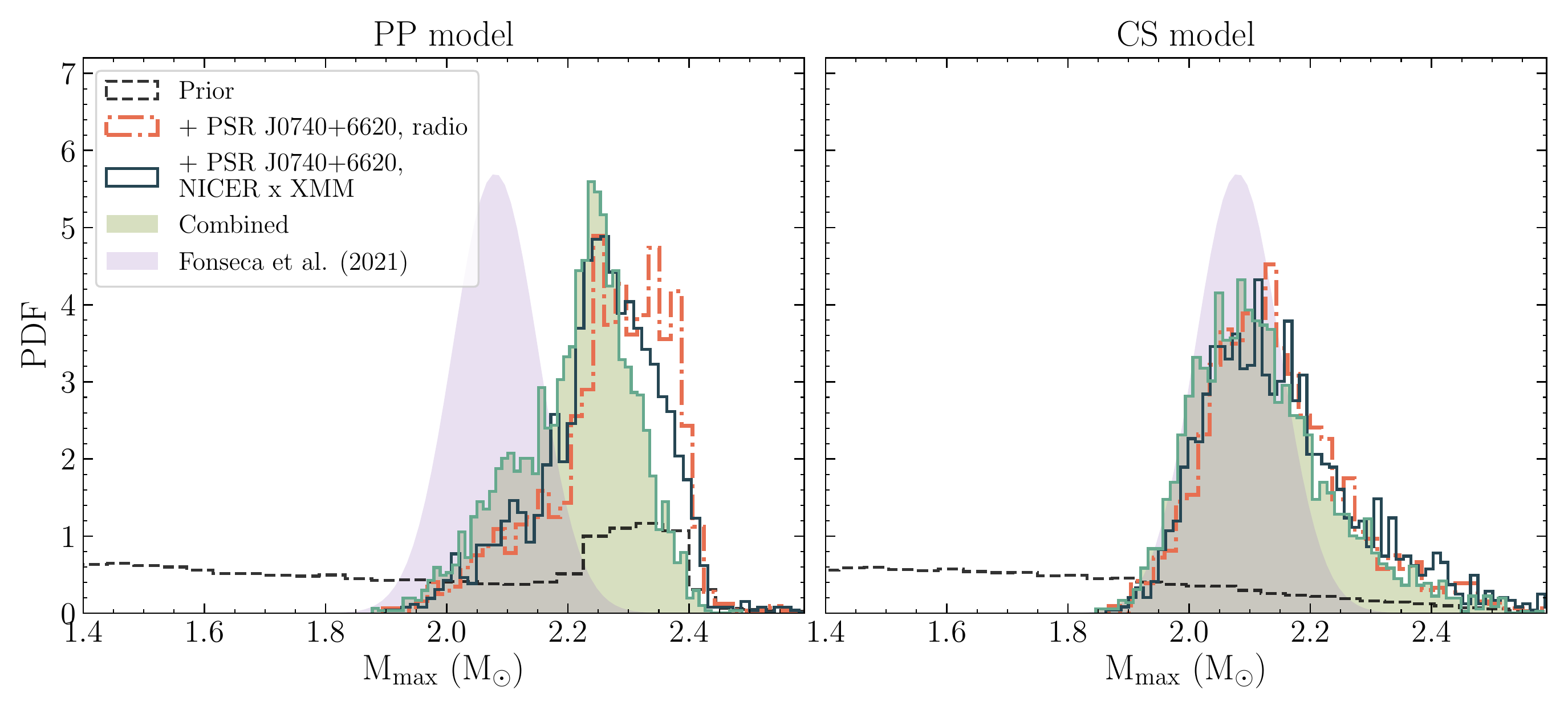}
\caption{Posterior distribution of the maximum mass of a non-rotating neutron star M$_{\rm TOV}$ for the PP model (left) and CS model (right) when considering only the radio mass measurement of \joh, the joint mass-radius estimate of \joh (\NICER $\times$ \textit{XMM}), and when combining \NICER's results on \joh and \jdbl with GW170817 and GW190425, and AT2017gfo. For the latter (``Combined'') we find a 95\% credible range for M$_{\rm TOV}=2.23^{+0.14}_{-0.23}\,M_\odot$ and M$_{\rm TOV}=2.11^{+0.29}_{-0.16}\,M_\odot$ for the PP and CS model, respectively. Also shown in pink is the radio mass measurement of \joh from \citet{Fonseca21}, as the heaviest pulsar measured to date.}
\label{fig:fig8}
\end{figure*}

An important quantity relating to the EOS is the maximum stable mass of a non-rotating neutron star, M$_{\rm TOV}$. Accurate knowledge of M$_{\rm TOV}$ can aid in classifying compact mergers and merger remnants. In Figure \ref{fig:fig8} we show posterior distributions on M$_{\rm TOV}$ when analyzing the updated radio mass measurement of \joh, the joint mass-radius estimate of \joh and combining GW170817, GW190425, AT2017gfo, \joh and \jdbl. The latter results in 95\% credible ranges for M$_{\rm TOV}=2.23^{+0.15}_{-0.25}$\,\msol and M$_{\rm TOV}=2.11^{+0.28}_{-0.16}$\,\msol for the PP and CS model, respectively. This is in agreement with values previous found \citep[see, e.g.,][and references therein]{Nathanail21} when assuming the secondary component in GW190814 was a black hole \citep{Abbott20_GW190814}. Note that the higher end of the distribution in Figure \ref{fig:fig8} is very dependent on our choice of parameterization, as no information is included from sources with masses above $2.08$\,\msol. One could use information on the merger remnant of GW170817 to put an upper bound on M$_{\rm TOV}$ \citep[see, e.g.,][]{Margalit17, Shibata17, Ruiz18}, but that is beyond the scope of this Letter. The lower end of the distribution on the other hand is strongly correlated with the radio mass measurement of \joh. The recently lowered mass distribution presented in \citet{Fonseca21} results in slightly lower values for M$_{\rm TOV}$ compared to the distributions found in \citet{Raaijmakers20}.

\subsection{Systematic uncertainties and framework comparisons}

The analysis presented in this Letter is conditional on both the modeling choices of the dense matter EOS and on modeling choices within each analysis of the multimessenger sources considered here. The sensitivity to the EOS modeling is explored here by employing two different high-density parameterizations and four different low-density chiral EFT calculations (see Section \ref{sec:ceft}). From Table \ref{tab:tab2} we conclude that the CS model systematically predicts lower radii, as a result of the additional constraints on the speed of sound that are not considered in the PP model. The discrepancy between the two models increases with increasing neutron star mass, as high-mass stars depend more sensitively on the choice of high-density parameterization. The two models considered here are however not exhaustive as many more high-density parameterizations exist \citep[see, e.g.,][]{Lindblom18,Capano:2019eae, Boyle20}. 

Furthermore, we do not consider the impact of any systematic effects present in estimating the posterior distributions on $M$, $R$ and $\Lambda$. For example, the uncertainty in modeling the hot regions in pulse profile modeling and the effect on the EOS has been studied in \citet{Raaijmakers20} by using two different models to fit \jdbl, which led to slightly different constraints. For \joh, different assumptions and priors lead to a higher estimated radius in the independent analysis of \cite{Miller21} (see the extensive discussion of this issue in Section 4.4 of \citealt{Riley21}), and we refer the reader to that paper for an EOS analysis using those results. \footnote{Note however that one of the main reasons for the higher inferred radius reported by \citet{Miller21} is that they do not truncate the prior on radius during the pulse profile modelling step, which \citet{Riley21} do (truncating above 16\,km, reflecting the lack of EOS models predicting higher radii, and thereby lowering the computational cost by reducing the parameter space). In the analysis by \citet{Miller21} the lack of prior support for high radii is effectively incorporated at a later stage, in the EOS analysis.}.

Measurements in $\Lambda$ from gravitational wave data are also sensitive to choice of priors and gravitational waveform models \citep[see, e.g.,][]{Kastaun:2019bxo, Gamba20}. Lastly, many different kilonova models exist \citep[see, e.g.,][for recent analyses]{Dietrich20, Nicholl21, Breschi21b} that derive slightly different constraints on the EOS due to differences in modeling assumptions on, e.g., geometry, composition and the connection between binary properties and outflow properties.

The inference framework employed in this Letter was first discussed in \citet{Riley18} and subsequently developed in \citet{Greif19, Raaijmakers19, Raaijmakers20}, which also introduced the chiral EFT constraints. Although an exhaustive comparison with other frameworks is out of the scope of this work, we will briefly mention similarities and differences with some commonly used frameworks in the field. Firstly, we make use of two particular high-density EOS parameterizations. Besides many different existing choices in these parameterizations, a completely different approach is to use non-parametric inference involving Gaussian Processes \citep[see, e.g.,][]{Landry19, Essick:2019ldf, Han20}, or discretely sampling a set of pre-computed EOS \citep[see, e.g.,][]{Capano:2019eae, Dietrich20}. Secondly, we compute likelihoods by performing kernel density estimation on posterior samples of neutron star properties such as mass, radius and tidal deformability \citep[see also, e.g,][]{Miller19, AlMamun21}. It is also possible to directly infer EOS properties from the observational data, for example X-ray or gravitational wave data. For the first, \citet{Riley18} argue that this approach would be computationally too expensive, while for the latter this has been done by, e.g., \citet{Capano:2019eae, Dietrich20}. A slightly different approach is used by \citet{HernandezVivanco20}, where the likelihood is computed by interpolating marginalized likelihoods using machine learning.

\subsection{Summary and future prospects}

In summary, the new joint mass-radius estimate of \joh significantly constrains the EOS. For the PP model the information gain is mostly a result of the high mass of the pulsar, as the $68\%$ credible range of the radius estimate exactly encompasses our prior distribution, informed by chiral EFT calculations, in that mass range. For the CS model the relatively high radius estimate does constrain the model at lower radii on top of constraints coming from the mass estimate. Combined with other current observational data from gravitational waves and kilonova light curves, as well as the \NICER mass-radius estimate of \jdbl, we find the 95\% credible ranges $12.38^{+0.70}_{-0.97}\,$km (PP model) and $12.23^{+0.48}_{-0.97}\,$km (CS model) for the radius of a $1.4$\,\msol neutron star. 

In the near future, the detailed analysis of gravitational wave events observed during the second part of the third observing run of LIGO/Virgo are expected to be published, among them a few candidate events which, in an initial rapid classification, were identified as containing at least one neutron star. 
 Any  measured tidal deformability from these gravitational waves events will help constrain the EOS further. There were unfortunately no EM counterparts for the potential binary neutron star or black hole-neutron star events during this observing run. The fourth observing run is planned to start next year, with the LIGO and Virgo detectors close to their design sensitivity and KAGRA fully joining the network~\citep{2020LRR....23....3A}. At design sensitivity, GW170817-like signals will have signal-to-noise ratios of 100 and enable measurements of tidal deformability with more than three times better accuracy~\citep{Capano:2019eae}. Subsequent further detector improvements are already planned for the mid to late 2020s~\citep{2020LRR....23....3A}, and an ongoing worldwide effort is paving the way for next decade's  third generation detectors. These will improve current measurements of tidal deformability by a factor of $\sim 10$ and observe the population of tens to hundreds of thousands of neutron star binaries, with EM counterparts detectable for a fraction of them \citep{Maggiore:2019uih,Sathyaprakash:2019yqt,Sathyaprakash:2019rom}.

In the coming months, \NICER is expected to deliver mass-radius measurements for three additional pulsars:  two for which independent mass constraints exist (the $\sim 1.4$\,\msol pulsar PSR J0437-4715 and the $\sim 1.9$\,\msol pulsar PSR J1614-2230); and the pulsar PSR J1231-1411, which has no independently known constraint on the mass. There will be an update to the inferred mass and radius of PSR J0030+0451, using a larger data set, taking into account improvements to our understanding of the \NICER instrument response, and including \textit{XMM} data in a joint analysis (as done for PSR J0740+6620). There are also good prospects for narrowing the mass-radius measurements for PSR J0740+6620, using models of the \NICER background. All of these promise further improvements to our understanding of the dense matter EOS. 

\section*{Acknowledgments}

This work was supported in part by NASA through the \NICER mission and the Astrophysics Explorers Program. G.R., T.H., S.N. and T.E.R. are grateful for support from the Nederlandse Organisatie voor Wetenschappelijk Onderzoek (NWO) through the VIDI and Projectruimte grants (PI Nissanke). T.H. also acknowledges support from the NWO sectorplan. T.E.R. and A.L.W. acknowledge support from ERC Consolidator Grant No.~865768 AEONS (PI: Watts). S.K.G., K.H., and A.S. acknowledge support from the Deutsche Forschungsgemeinschaft (DFG, German Research Foundation) -- Project-ID 279384907 -- SFB 1245. W.C.G.H. acknowledges support through grant 80NSSC20K0278 from NASA. This work was sponsored by NWO Exact and Natural Sciences for the use of supercomputer facilities, and was carried out on the Dutch national e-infrastructure with the support of SURF Cooperative. We thank Sharon Morsink and Renee Ludlam for comments on a draft manuscript. We thank Mansi Kasliwal for providing us with the data for AT2017gfo. We thank the anonynomous referee for providing useful feedback on the manuscript.

\software{Python/C~language~\citep{python2007}, GNU~Scientific~Library~\citep[GSL;][]{Gough:2009}, NumPy~\citep{Numpy2011}, Cython~\citep{cython2011}, SciPy~\citep{Scipy}, MPI~\citep{MPI}, \project{MPI for Python}~\citep{mpi4py}, Matplotlib~\citep{Hunter:2007,matplotlibv2}, IPython~\citep{IPython2007}, Jupyter~\citep{Kluyver:2016aa}, \MultiNest~\citep{Feroz09}, \textsc{PyMultiNest}~\citep{Buchner14}, kalepy~\citep{Kelley2021}.} 

\bibliography{References}

\end{document}